# Cryogenic GaAs high-electron-mobility-transistor amplifier for current noise measurements

Sanghyun Lee[1,2], Masayuki Hashisaka[1,3,a)], Takafumi Akiho[1], Kensuke Kobayashi[2], and Koji Muraki[1]

**AFFILIATIONS**

[1]NTT Basic Research Laboratories, NTT Corporation, 3-1 Morinosato-Wakamiya, Atsugi, Kanagawa 243-0198, Japan
[2]Graduate School of Science, Osaka University, 1-1, Machikaneyama, Toyonaka, Osaka 560-0043, Japan
[3]JST, PRESTO, 4-1-8 Honcho, Kawaguchi, Saitama 332-0012, Japan

a)Author to whom correspondence should be addressed: masayuki.hashisaka.wf@hco.ntt.co.jp

**ABSTRACT**

We show that a cryogenic amplifier composed of a homemade GaAs high-electron-mobility transistor (HEMT) is suitable for current-noise measurements in a mesoscopic device at dilution-refrigerator temperatures. The lower noise characteristics of our homemade HEMT leads to a lower noise floor in the experimental setup and enables more efficient current-noise measurement than is available with a commercial HEMT. We present the dc transport properties of the HEMT and the gain and noise characteristics of the amplifier. With the amplifier employed for current-noise measurements in a quantum point contact, we demonstrate the high resolution of the measurement setup by comparing it with that of the conventional one using a commercial HEMT.

## I. INTRODUCTION

Mesoscopic systems have served as important experimental platforms for studying the quantum nature of electrons and associated correlated effects. Various quantum effects have been investigated by measuring transport properties of the systems. Conductance $G$ is often evaluated by measuring dc current that corresponds to the time average of the number of electrons passing through a device. In addition, current noise, which corresponds to the fluctuation of the number of electrons, is useful for gaining deeper insights into quantum transport[1-3]. For example, noise measurements have successfully revealed fractional charge of tunneling quasiparticles in fractional quantum Hall systems[4-7] and correlated electron transport through Kondo impurities[8-10].

Despite its high potential, current-noise measurement has not been widely performed because of its technical difficulties. The central problem is that the power spectral density (PSD) of current noise, $S^I \equiv \langle \Delta I^2 \rangle$, in a mesoscopic device is too small (typically below $10^{-28}$ A$^2$/Hz) to measure with a standard ammeter. In previous experiments, cryogenic low-noise amplifiers[4-25] have been employed to solve this problem. For example, it was reported that homemade amplifiers consisting of a commercial high-electron-mobility transistor (HEMT) (Avago Technologies ATF-34143) operating at 4.2 K provides resolution of $\delta S^I_{in} \cong 2.8 \times 10^{-29}$ A$^2$/Hz in a cross-correlation current-noise measurement with a 10-s data integration time $\tau_{int}$[11]. While this value is already sufficient in many cases, further improvement of the resolution is desirable to probe novel mesoscopic phenomena such as anyonic correlations in fractional quantum Hall systems[26,27] and violation of Bell inequalities in an electronic interferometer[28].

In this paper, we report a cryogenic common-source (CS) amplifier composed of a homemade GaAs HEMT. The GaAs/AlGaAs heterostructure and gate pattern of the HEMT were designed to attain high transconductance and hence low-noise characteristics. We calibrated the measurement system through Johnson noise thermometry and measured the shot noise generated at a quantum point contact (QPC) at a dilution-refrigerator temperature. The resolution in an auto-correlation measurement with $\tau_{int} = 50$ s is $\delta S^I_{in} \cong 0.52 \times 10^{-29}$ A$^2$/Hz, corresponding to less than one third of that reported for a similar system using a commercial-HEMT amplifier[11]. Because the statistical error decreases in inverse proportion to $\sqrt{\tau_{int}}$, our system improves the efficiency of current-noise measurement by more than a factor of about $(1/3)^{-2} = 9$.

This paper is organized as follows. Section II presents an overview of our noise measurement system. Section III describes our GaAs HEMTs. Section IV discusses the cryogenic CS amplifiers. Section V demonstrates the current-noise measurements performed on a QPC. Section VI summarizes the paper.

## II. NOISE-MEASUREMENT SYSTEM

### A. Overview

We installed a cryogenic CS amplifier in a standard measurement setup comprising an inductor-capacitor (LC) tank circuit[11-13]. Figure 1(a) shows a block diagram of our setup installed in a dilution refrigerator, and Fig. 1(b) illustrates a circuit model of the cryogenic assembly. A bias voltage $V_{bias}$ is applied to generate a current $I_m$ flowing into a mesoscopic device (resistance $R_m$) at the mixing-chamber (MC) temperature. The transmitted current $I_{in}$ flows down to the cold ground through $L_{in} = 33$ μH placed on the MC plate. Note that, even if $I_m$ is noiseless, the current flowing through the device may reflect the discrete nature of electron charge, leading to the fluctuation $\Delta I_{in}$ in $I_{in}$. The inductor forms the tank circuit with capacitance $C_{in}$, mainly composed of the parasitic capacitance of the coaxial cable, $C_{coax1}$. Near the LC resonance frequency, $\Delta I_{in}$ yields voltage noise $\Delta V_{in} = Z_1 \times \Delta I_{in}$, where

$$Z_1 = \left[\frac{1}{R_m} + (i\omega L_{in} + r)^{-1} + i\omega C_{in}\right]^{-1} \quad (1)$$

is the parallel impedance of the mesoscopic device and the tank circuit. Here, $\omega$ is the angular frequency and $r$ is the total parasitic resistance of the inductor and the coaxial cable. The voltage noise is amplified to $\Delta V_{out}$ by the cryogenic amplifier and again amplified to $\Delta V_{meas}$ by a commercial amplifier (NF Corporation SA-220F5, 46 dB) at room temperature. The measurement is completed by recording $\Delta V_{meas}$ using a digitizer (National Instruments PXI-5922) that serves as a high-speed voltmeter. We analyze the time-domain data by the fast Fourier transform (FFT) technique to evaluate $S^I_{in} \equiv \langle \Delta I_{in}^2 \rangle$ from the FFT spectrum $S_{meas} \equiv \langle \Delta V_{meas}^2 \rangle$.

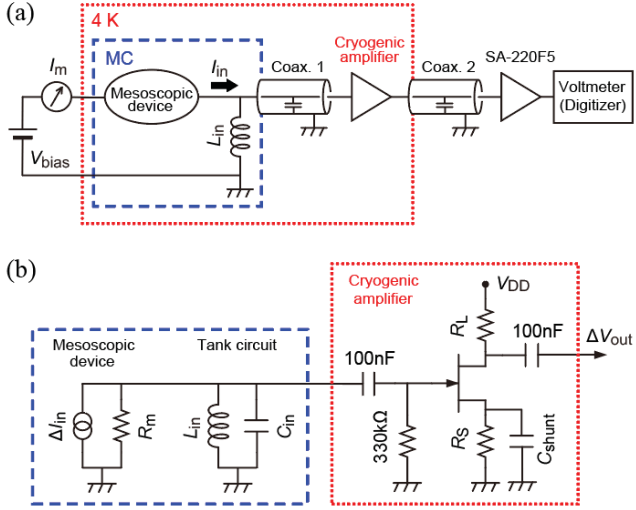

Fig. 1. (a) Overview of the noise-measurement setup using an LC tank circuit. Mesoscopic device and amplifiers are connected via CuNi coaxial cables, Coax. 1 and Coax. 2 with parasitic capacitances $C_{coax1}$ and $C_{coax2}$, respectively. Cryogenic amplifier is placed on the 4-K stage of the dilution refrigerator. (b) Electronic circuit model of the cryogenic assembly.

In this study, we focus on the Johnson noise and the shot noise in the low-frequency white-noise limit ($f \ll k_B T_e/h$ and $eV_{bias}/h$, where $e$ is the elementary charge, $h$ is the Planck constant, $k_B$ is the Boltzmann constant, and $T_e$ is electron temperature). At very low frequencies (typically below 100 kHz), however, these noises are buried in the $1/f$ noise generated in the mesoscopic device and/or the cryogenic HEMT amplifier. To avoid the $1/f$ noise, we designed the LC resonance frequency $f_1 = (2\pi\sqrt{L_{in}C_{in}})^{-1}$ to be near 1.8 MHz, where the $1/f$ noise is expected to be small, by choosing $L_{in} = 33$ μH and $C_{in} \cong C_{coax1} \cong 240$ pF.

**B. Common-source circuit**

Here, we briefly review the characteristics of a CS circuit, CS1, shown in Fig. 2, with a gate supply voltage $V_g$ and a drain voltage $V_{DD}$. The 330-kΩ resistor and the 100-nF capacitor connected to the HEMT gate forms an RC filter that attenuates high-frequency extrinsic noise from the gate supply. For a set of $V_g$ and $V_{DD}$, the gain

$$A_{CS1}(V_g, V_{DD}) \equiv \frac{\partial V_{out}(V_g, V_{DD})}{\partial V_g} \quad (2)$$

is given by

$$A_{CS1} = -g_m \frac{R_L}{1+R_L g_{ds}} = -g_m Z_{out}, \quad (3)$$

where

$$g_m(V_g, V_{ds}) \equiv \frac{\partial I_{ch}(V_g, V_{ds})}{\partial V_g}, \quad (4)$$

and

$$g_{ds}(V_g, V_{ds}) \equiv \frac{\partial I_{ch}(V_g, V_{ds})}{\partial V_{ds}}. \quad (5)$$

This leads to

$$A_{CS1} = -R_L \frac{\partial I_{ch}(V_g, V_{DD})}{\partial V_g}. \quad (6)$$

Here, $g_m$ is the transconductance, $g_{ds}$ is the drain conductance, $R_L$ is the load resistance, $Z_{out} = R_L/(1+R_L g_{ds})$ is the output impedance of the amplifier, $V_{ds} = V_{DD} - R_L I_{ch}$ is the voltage between the drain and source of the HEMT, and $I_{ch}(V_g, V_{ds})$ is the current flowing in the HEMT channel.

**C. Noise in a CS amplifier**

The signal-to-noise ratio in the first amplification step, namely the cryogenic part in Fig. 1(a), governs the resolution of the measurement system. While the cryogenic amplifier converts $S^I_{in}$ to $S^V_{out} \equiv \langle \Delta V_{out}^2 \rangle$ in conjunction with the LC tank circuit, it also generates extrinsic noise, which interferes with $S^I_{in}$. With this extrinsic noise included, the relation between $S^I_{in}$ and $S^V_{out}$ can be described as

$$S^V_{out} = |A(f)|^2 [|Z_1^2|(S^I_{in} + S^I_{HEMT}) + S^V_{HEMT}]. \quad (7)$$

Here, $A(f)$ is the gain of the cryogenic amplifier, and $S^I_{HEMT}$ and $S^V_{HEMT}$ are the input-referred current and voltage noise of the HEMT, respectively. When the gate leakage current of the HEMT is negligible, both $S^I_{HEMT}$ and $S^V_{HEMT}$ originate exclusively from the current noise $S^I_{ch} \equiv \langle \Delta I_{ch}^2 \rangle$ in the HEMT channel generated by the finite source-drain

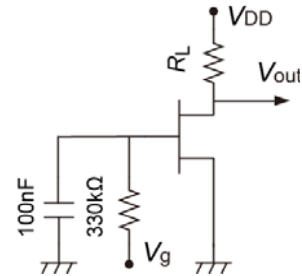

Fig. 2. Common-source circuit CS1.

voltage[29].

At cryogenic temperatures, $S^I_{ch}$ is given by[23]

$$S^I_{ch} \cong S^I_{ch-1/f} + S^I_{ch-shot}, \quad (8)$$

where $S^I_{ch-1/f}$ and $S^I_{ch-shot}$ are the PSDs of the 1/f noise and the shot noise generated in the HEMT channel, respectively. It is empirically known that $S^I_{ch-1/f}$ increases in proportion to $I_{ch}^2$ and decreases in inverse proportion to the total number of charge carriers[30,31]. On the other hand, $S^I_{ch-shot}$ is proportional to $I_{ch}$ as

$$S^I_{ch-shot} = 2eI_{ch}F, \quad (9)$$

where $F$ is the so-called Fano factor ($0 \leq F \leq 1$).

## III. HOMEMADE HEMT

In principle, $S^V_{HEMT}$ can be suppressed by increasing $g_m \equiv \Delta I_{ch}/\Delta V_g$ while keeping $S^I_{ch}$ low. For this purpose, we fabricated HEMTs using a GaAs/AlGaAs heterostructure and gate patterns that are designed to be suitable for high $g_m$. Transport properties of the homemade HEMTs were measured at temperatures below 4.2 K, where these HEMTs show no significant temperature dependence. We note that the major features of the transport properties are unchanged after several cool downs.

### A. Heterostructure

Figure 3(a) shows a schematic of the HEMT fabricated from a GaAs/Al$_{0.33}$Ga$_{0.67}$As heterostructure grown by molecular beam epitaxy on a semi-insulating GaAs substrate. The heterostructure is modulation-doped with silicon at two δ planes (doping level of $6 \times 10^{12}$ cm$^{-2}$ for each). The two-dimensional electron system (2DES) located 55 nm below the surface has electron density $n_e = 4.0 \times 10^{11}$ cm$^{-2}$ and mobility $\mu = 3.2 \times 10^5$ cm$^2$V$^{-1}$s$^{-1}$ at 4.2 K. The shallow depth and the high electron density of the 2DES are advantageous for increasing $g_m$.

### B. Device fabrication

Our HEMT was patterned by photolithography for fabricating mesa structures, ohmic contacts of Au-Ge-Ni alloys, and a gate electrode of 10-nm-thick titanium and 30-nm-thick gold.

For a given GaAs/AlGaAs heterostructure, the HEMT characteristics are determined by the geometry of the gated region: gate length $L$ and channel width $W$. To examine how $g_m$ and the noise characteristics depend on $L$ and $W$, we fabricated five HEMTs with different sets ($W$; $L$) on the same wafer. Four of them have ($W$; $L$) = (1 mm; 2, 4, 16, or 64 μm), and the other has (3 mm; 4 μm) [see Fig. 3(b) for an example].

### C. dc transport properties

Figure 4(a) presents $V_g$-$I_{ch}$ traces of the four HEMTs with different $L$, measured at $V_{ds} = 0.27$ V below 1 K. For a shorter $L$, a more negative $V_g$ is necessary to pinch-off the channel. Concomitantly, the slope of the traces becomes steeper for a shorter $L$, resulting in a higher $g_m$. Figure 4(b) displays the $V_g$ dependence of $g_m$ of the HEMTs with $L = 2$ μm and 4 μm, obtained by numerically differentiating $I_{ch}$ with respect to $V_g$. While the 4-μm HEMT shows a single gentle peak of height $g_m \cong 80$ mS, the 2-μm one shows a double peak reflecting the irregular structure in the pinch-off trace [indicated by the vertical arrow in Fig. 4(a)]. The double-peak structure is sensitive to a slight change in $V_{ds}$, which results in HEMT instability. Similar irregular features were observed in several $L = 2$ μm HEMTs, while the details differ from one another. We consider that the irregularities originate from unintentional tunneling through impurities or defects in the gated region. Note that a higher $V_{ds}$ sometimes induces similar irregularities even in the HEMTs with longer $L$. However, for longer $L$ such irregularity appears only at $V_{ds}$ much higher than in the $L = 2$ μm HEMT. Thus, we conclude that the $L = 4$ μm HEMT is best suited for the present study because of its high stability and high $g_m$.

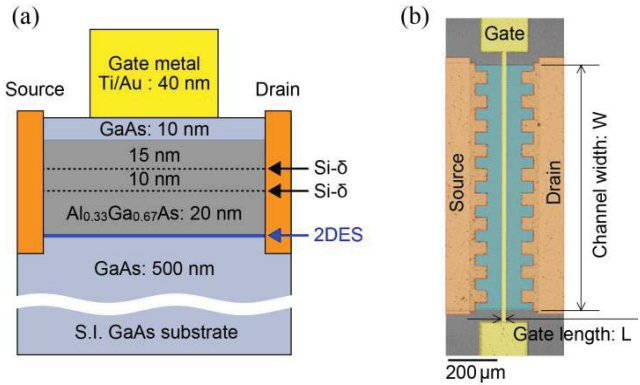

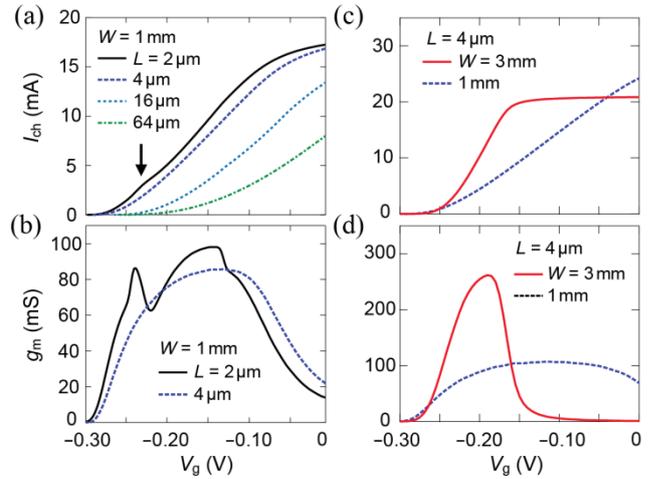

Fig. 3. (a) Schematic of the HEMT structure fabricated in 2DES in GaAs/AlGaAs heterostructure. (b) False-color optical micrograph of a ($W$; $L$) = (1 mm; 16 μm) HEMT. The meander structure of source and drain electrodes is for suppressing ohmic contact resistances.

Fig. 4. (a) $V_g$ dependence of $I_{ch}$ of the four HEMTs with different $L$. $V_{ds} = 0.27$ V. (b) $V_g$ dependence of $g_m$ of HEMTs with $L = 2$ μm and $L = 4$ μm. (c) $V_g$ dependence of $I_{ch}$ of HEMTs with $W = 3$ mm and $W = 1$ mm. $V_{ds} = 0.5$ V. (d) $g_m$ corresponding to the case shown in (c).

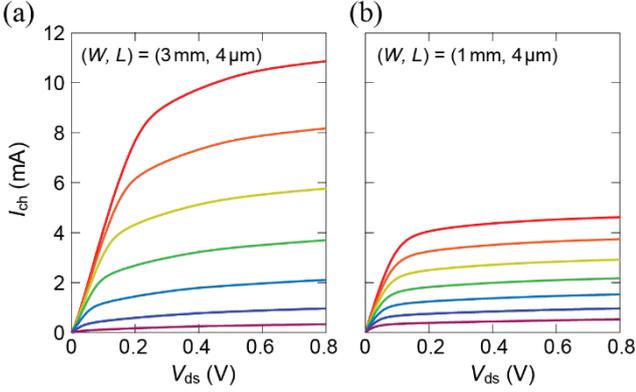

Fig. 5. (a) $V_{ds}$-$I_{ch}$ traces of $(W; L) = (3\text{ mm}; 4\text{ μm})$ HEMT measured at several $V_g$ values from −0.20 V (red) to −0.26 V (purple) in 0.01 V steps. (b) $V_{ds}$-$I_{ch}$ traces of $(W; L) = (1\text{ mm}; 4\text{ μm})$ HEMT in the same range of $V_g$.

Figure 4(c) shows the pinch-off characteristics of the (3 mm; 4 μm) and (1 mm; 4 μm) HEMTs measured at $V_{ds} = 0.5$ V at 4.2 K[32]. While the pinch-off voltages are similar in the two devices, the change in $I_{ch}$ is much steeper in the $W = 3$ mm HEMT at $-0.25\text{ V} < V_g < -0.18\text{ V}$. In the $g_m$-$V_g$ plot shown in Fig. 4(d), $g_m$ of the $W = 3$ mm HEMT reaches ~260 mS at $V_g = -0.19$ V, which is about 2.5 times that of the $W = 1$ mm one.

Figures 5(a) and 5(b) show the $V_{ds}$ dependence of $I_{ch}$ of the (3 mm; 4 μm) and (1 mm; 4 μm) HEMTs measured at several $V_g$ values between −0.20 and −0.26 V. The drain conductance at a given set of ($V_g$, $V_{ds}$) can be evaluated by differentiating $I_{ch}$ with respect to $V_{ds}$; for example, at ($V_g$, $V_{ds}$) = (−0.25 V, 0.5 V), we find $g_{ds} \cong 0.52$ mS for the $W = 3$ mm HEMT and 0.24 mS for the $W = 1$ mm one.

### D. Noise characteristics

We measured the noise characteristics of the (3 mm; 4 μm) and (1 mm; 4 μm) HEMTs and a commercial HEMT (Avago Technologies ATF-35143, referred to as the "ATF HEMT" below) in the CS1 circuit shown in Fig. 2 at 4.2 K. We chose $R_L = 500$ Ω, where $Z_{out} \cong R_L = 500$ Ω. In this case, $S^I_{ch}$ and $S^V_{HEMT}$ can be evaluated as

$$S^I_{ch} = S^V_{out}/|Z_{out}|^2 \cong S^V_{out}/R_L^2, \quad (10)$$

$$S^V_{HEMT} = S^V_{out}/|A_{CS1}|^2. \quad (11)$$

Before examining the noise characteristics, we measured the dc response of the CS1 circuits to choose the operating point. Figure 6(a) shows the $V_g$ dependence of $I_{ch}$ (upper panel) and $|A_{CS1}|$ (lower panel) of the GaAs-HEMT circuits at $V_{DD} = 1$ V. Figure 6(b) show that of the ATF HEMT measured at $V_{DD} = 0.4$ V[33]. When $V_g$ is decreased from 0 V, we observe $I_{ch} \cong V_{DD}/R_L$ down to a threshold $V_g$ value [e.g., −0.24 V for the (3 mm; 4 μm) HEMT], where the resistance of the HEMT channel becomes comparable to $R_L$. When $V_g$ is further decreased, $I_{ch}$ decreases to zero, with the slope corresponding to $|A_{CS1}|$ on the way. Here, we chose the $|A_{CS1}|$ peak as the operating point. For example, the operating point of the (3 mm; 4 μm) HEMT was set at $V_g = -0.26$ V, where $|A_{CS1}| \cong 32$, at $V_{DD} = 1$ V.

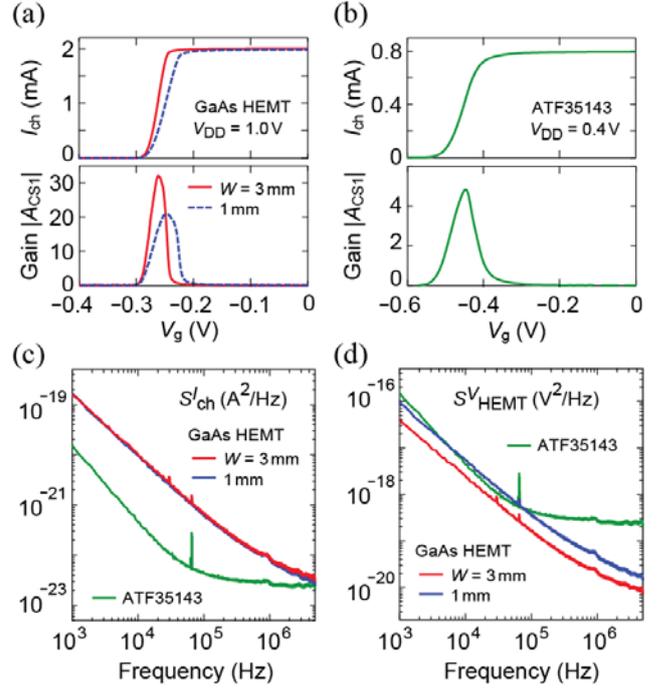

Fig. 6. (a) $V_g$ dependence of $I_{ch}$ (upper) and $|A_{CS1}|$ (lower panel) of (1 mm; 4 μm) and (3 mm; 4 μm) HEMT circuits at $V_{DD} = 1.0$ V. (b) $I_{ch}$ and $|A_{CS1}|$ traces of the ATF-HEMT circuit at $V_{DD} = 0.4$ V. (c) $S^I_{ch}$ and (d) $S^V_{HEMT}$ of the three CS1 circuits.

Figures 6(c) and 6(d) respectively show $S^I_{ch}$ and $S^V_{HEMT}$ spectra at the operating points, estimated from the measured $S^V_{out}$ spectra using Eqs. (10) and (11). In these plots, the RC damping at the output due to $Z_{out} \cong R_L = 500$ Ω and $C_{coax2} \cong 75$ pF [see Fig. 1(a)] are numerically compensated. Note that the data for the GaAs HEMTs were obtained at $V_{DD} = 1$ V, while that for the ATF one was obtained at 0.4 V. At low frequencies, where the $1/f$ noise governs the noise characteristics, $S^I_{ch}$ of the GaAs HEMTs is much larger than that of the ATF HEMT, mainly because $V_{DD}$ is larger in the former. Meanwhile, near the noise-measurement frequency of $f_1 \cong 1.8$ MHz (see Sec. II), where the shot-noise contribution becomes dominant, the $S^I_{ch}$ values of the three HEMTs are comparable to each other. The low-noise performance of the GaAs HEMTs manifests itself in the $S^V_{HEMT}$ spectra; the higher gains of the GaAs HEMTs result in the lower $S^V_{HEMT}$ [see Figs. 6(a) and 6(b) and Eq. (11)]. Actually, among the three, the (3 mm; 4 μm) HEMT circuit, which has the highest gain, has the lowest $S^V_{HEMT}$ over the entire frequency range. We note that near $f_1 \cong 1.8$ MHz, it is about one order of magnitude lower than that of the ATF one [Fig. 6(d)].

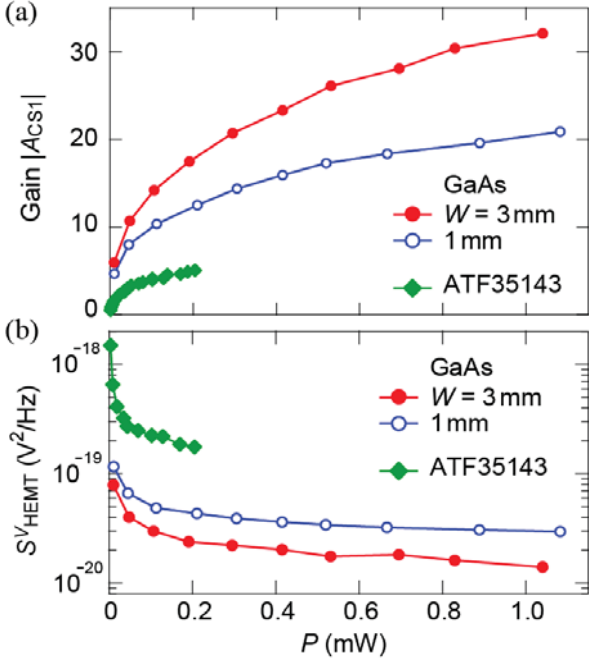

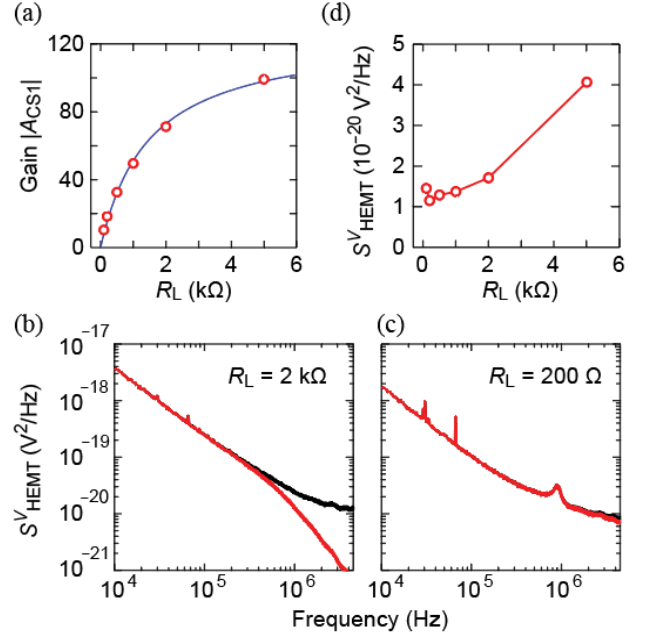

Fig. 7. (a) Gain $|A_{CS1}|$ of the three CS1 circuits plotted as a function of $P$. (b) $P$ dependence of $S^V_{HEMT}$ at $f = 1.8$ MHz.

Fig. 8. (a) Gain $|A_{CS1}|$ as a function of $R_L$ at $P \cong 1$ mW. Blue line is a simulated curve using Eq. (3). (b)(c) $S^V_{HEMT}$ spectra of CS1 circuits with (b) $R_L = 2$ kΩ and (c) 200 Ω. Black (red) curves are with (without) compensation of RC damping[35]. (d) $R_L$ dependence of $S^V_{HEMT}$ at $f = 1.8$ MHz. The RC damping is numerically compensated.

The measured $|A_{CS1}|$ and $S^V_{HEMT}$ are summarized in Figs. 7(a) and 7(b), respectively, as a function of power consumption $P = V_{DD}I_{ch}$. The (3 mm; 4 μm) HEMT shows the highest $|A_{CS1}|$ and lowest $S^V_{HEMT}$ over the entire range of $P$[34] and is therefore best suited for the present purpose. The gain, $|A_{CS1}| \propto \partial I_{ch}/\partial V_g$, monotonically increases with $P$, reflecting the $\sqrt{P}$ dependence of $I_{ch}$. On the other hand, $S^V_{HEMT}$ decreases with increasing $P$ due to the suppression of the shot-noise contribution $S^I_{ch\text{-}shot} \times R_L^2/|A_{CS1}|^2 \propto P^{-1/2}$ [see Eqs. (8-11)].

## IV. CRYOGENIC GaAs-HEMT AMPLIFIER

In this section, we discuss CS amplifiers based on the (3 mm; 4 μm) HEMT, which has the highest $|A_{CS1}|$ and the lowest $S^V_{HEMT}$.

### A. Load resistance

The load resistance $R_L$ determines $A_{CS1}$, as seen in Eqs. (3) and (6). In Fig. 8(a), the red circles show the measured $|A_{CS1}|$ for several $R_L$ values at $P \cong 1$ mW at 4.2 K. The gain increases monotonically with $R_L$, following the curve simulated using Eq. (3) with $g_m = 87$ mS and $g_{ds} = 0.67$ mS obtained from the fit to the data. The tiny deviations of the experimental data from the simulation are due to the changes in $g_m$ and $g_{ds}$ caused by the shift of the operating point depending on $R_L$.

The black (red) trace in Fig. 8(b) shows the representative $S^V_{HEMT}$ spectrum for $R_L = 2$ kΩ with (without) the numerical compensation for the RC damping at the output of CS1. Figure 8(c) shows those for $R_L = 200$ Ω. The effect of the RC damping in the $R_L = 200$ Ω circuit is much smaller than that in the $R_L = 2$ kΩ one, because $Z_{out}$, forming an RC filter with $C_{coax2}$ [see Fig. 1(a)], monotonically decreases with $R_L$. Figure 8(c) indicates that below $R_L = 200$ Ω the damping effect can be neglected at the noise-measurement frequency of $f_1 \cong 1.8$ MHz.

Figure 8(d) shows the $R_L$ dependence of $S^V_{HEMT}$ at 1.8 MHz. The $S^V_{HEMT}$ value is minimal near $R_L = 200$ Ω. Because of the negligible RC damping and the low $S^V_{HEMT}$, we chose $R_L = 200$ Ω for our cryogenic amplifier[36].

### B. Self-biasing

Unlike the CS1 circuit, the cryogenic amplifier in Fig. 1(b) has a self-biasing resistor $R_S$ and a shunt capacitor $C_{shunt}$. We examined the self-biasing using another common-source circuit, CS2, shown in Fig. 9. At megahertz frequencies, $C_{shunt}$ = 100 nF dominates over $R_S$ to ground the HEMT source, enabling us to obtain high ac gain $|A_{CS2}| = V_{out}/V_{ext}$ while simplifying the cryogenic assembly by removing the wiring for $V_g$ (see Fig. 2). We chose $R_S$ = 165 Ω and applied $V_{DD}$ = 0.883 V to set the circuit near the operating point, where $P \cong 1.3$ mW. Figure 10(a) shows an $|A_{CS2}|$ spectrum obtained by sweeping the frequency of the external ac voltage of $V_{ext}$ = 1 mV RMS and measuring the ac output. Thanks to $C_{shunt}$, $|A_{CS2}|$ increases with frequency to saturate at $|A_{CS2}| \cong 15.6$ above a few hundred kilohertz, which is comparable to the gain of the CS1 circuit $|A_{CS1}| \cong 18$ [see Fig. 8(a)]. When the frequency is increased above 2 MHz, $|A_{CS2}|$ decreases because of the RC damping at the output of the amplifier.

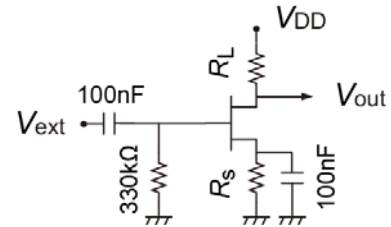

Fig. 9. Common-source circuit CS2.

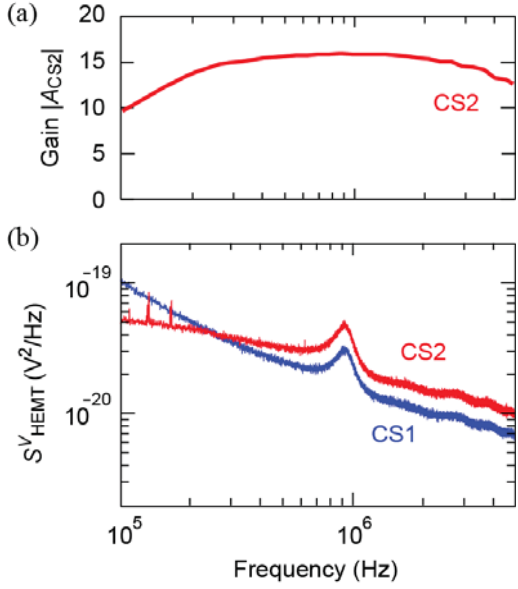

Fig. 10. (a) Measured $|A_{CS2}|$ spectrum. (b) Comparison of $S^V_{HEMT}$ between CS1 and CS2[35].

Figure 10(b) shows that, near 1.8 MHz, $S^V_{HEMT}$ of CS2 ($\cong 1.53 \times 10^{-20}$ A²/Hz) is slightly larger than that of CS1, probably because of the misalignment of the operating point. However, it still remains much smaller than that of the ATF one [see Figs. 6(d) and 7(b)].

## V. NOISE MEASUREMENT FOR A QPC

Here, we demonstrate current-noise measurements performed on a QPC using the system shown in Fig. 1(a). The cryogenic amplifier with $R_L$ = 200 Ω and $R_S$ = 165 Ω was activated by applying $V_{DD}$ = 0.883 V, i.e., using the same set of parameters ($R_L$, $R_S$, and $V_{DD}$) as those of the CS2 circuit presented in Sec. IV. While the amplifier shows a slight change in gain for different cool downs, its influence can be eliminated after appropriate calibration as described in this section. The measurement results are compared with those of the ATF-HEMT amplifier, which has the same $R_L$ and $R_S$ values and is activated by $V_{DD}$ = 1.9 V. The measurements were performed at 30 mK at zero magnetic field.

### A. Measurement setup

Figure 11(a) shows a schematic of the measurement setup. The QPC, fabricated in a 2DES with electron density $n_e = 1.5 \times 10^{11}$ cm⁻² and mobility $\mu = 6.6 \times 10^5$ cm²V⁻¹s⁻¹ in a GaAs/Al$_{0.33}$Ga$_{0.67}$As heterostructure, was formed by applying a gate bias $V_{SG}$. We measured the dc transport properties using a standard lock-in technique by applying an ac modulation of $V_{bias}$ = 10 µV RMS (33 Hz) and measuring the current $I_m$.

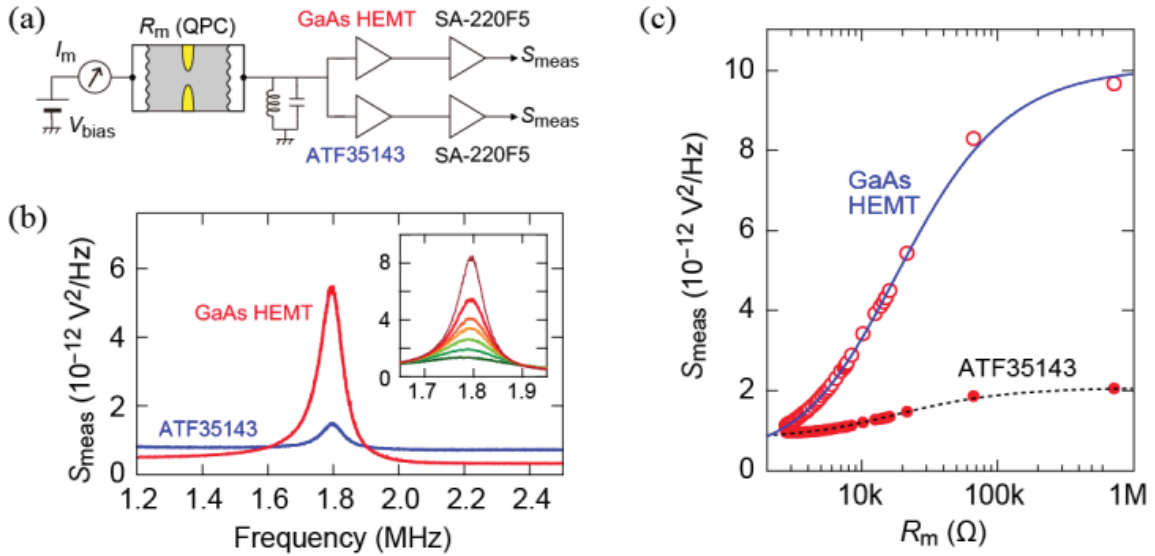

Fig. 11. (a) Schematic of noise-measurement setup for a QPC. (b) Representative noise spectra measured through GaAs line and ATF line. (Inset) Zero-bias noise spectra measured through GaAs line at $R_m$ = 66.9 kΩ (dark red), 21.5 kΩ (red), 13.7 kΩ (dark orange), 10.2 kΩ (orange), 7.7 kΩ (light green), 5.3 kΩ (green), and 3.6 kΩ (dark green). (c) $R_m$ dependence of peak heights for GaAs line (open circles) and ATF line (filled circles). Blue solid and black dashed lines show fitted curves.

The current noise was measured through the two measurement lines connected to a single output of the QPC: one containing the GaAs-HEMT amplifier ("GaAs line") and the other the ATF-HEMT amplifier ("ATF line"). We measured time-domain $\Delta V_{meas}$ data for $\tau_{int}$ = 50 s at a sampling rate of 10 MS/s and evaluated $S_{meas}$ spectra near 1.8 MHz. Figure 11(b) shows representative results obtained at $V_{SG}$ = –0.88 V, where the QPC resistance is $R_m$ = 21.5 kΩ. The resonance-peak height at $f_1$ = 1.794 MHz of the GaAs line is much higher than that of the ATF line, while the background of the former is lower than the latter. This observation suggests that the GaAs line has better resolution than the ATF line.

The resonance line shapes are slightly distorted from the Lorentzian line shapes expected for an ideal RLC resonance circuit. This distortion is due to the parasitic resistance $r$ in the RLC tank circuit and the parasitic capacitances in the HEMT.

**B. Calibration**

At $V_{bias}$ = 0 V, the Johnson noise dominates over other noises in the QPC, leading to $S^I_{in} \cong 4k_B T_e \mathrm{Re}(Y_1)$, where $\mathrm{Re}(Y_1)$ is the real part of the admittance $Y_1 = Z_1^{-1}$. Here, we calibrated the measurement system using the resonance peak height of the Johnson-noise spectra. The $S_{meas}$ peak height is described as

$$S_{meas} = A_{RT}^2 |A(f_1)|^2 [|Z_1^2|(S^I_{in} + S^I_{HEMT}) + S^V_{HEMT}], \quad (12)$$

where $A_{RT}$ (= 400) is the gain of the room-temperature amplifier [SA-220F5 in Fig. 1(a)] and $|A(f_1)|$ is that of the cryogenic amplifier at $f = f_1$ [see Eq. (7)]. With increasing $R_m$ by squeezing the QPC, $|Z_1|$ and hence the peak height monotonically increase, as shown in the inset of Fig. 11(b). Open (filled) circles in Fig. 11 (c) summarize the $R_m$ dependence of the $S_{meas}$ peak value measured through the GaAs (ATF) line, with a 30-kHz bandwidth around $f = f_1$. The blue solid (black dashed) curve is the simulation using Eq. (12)[37]. Table 1 summarizes the parameters used for the simulations. Here, we note two important observations that justify our analysis. First, the same $T_e$ and $S^I_{HEMT}$ values give a good agreement with the data for both GaAs and ATF lines over three orders of magnitude variation in $R_m$. Second, $|A(f_1)|$ and $S^V_{HEMT}$ of both amplifiers are comparable to the results obtained in Sec. III and Sec. IV.

| | $|A(f_1)|$ | $T_e$ | $S^I_{HEMT}$ | $S^V_{HEMT}$ |
|---|---|---|---|---|
| GaAs | 14.9 | 80 mK | 2.64 | 0.96 |
| ATF | 5.35 | 80 mK | 2.64 | 17.8 |

Table 1. Parameters of the fit curves in Fig. 11(c). The units of $S^I_{HEMT}$ and $S^V_{HEMT}$ are $10^{-27}$ A²/Hz and $10^{-20}$ V²/Hz, respectively.

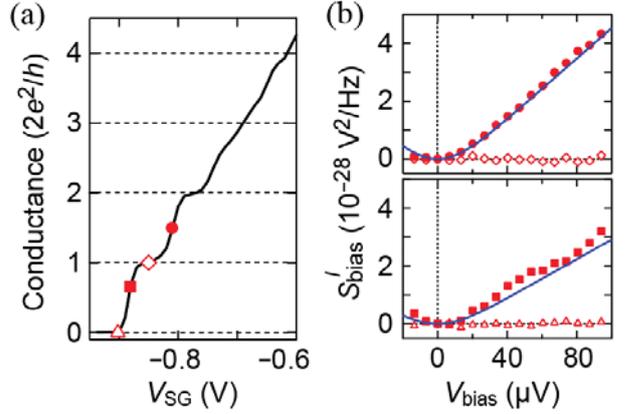

Fig. 12. (a) Conductance behavior of the QPC as a function of $V_{SG}$. (b) $V_{bias}$ dependence of $S^I_{bias}$ measured at $G/G_0$ = 1.5 (filled circles), 1 (open diamonds), 0.65 (filled squares), and 0 (open triangles). Blue solid curves are simulations using Eq. (13).

**C. Shot-noise measurements**

We measured the $V_{bias}$ dependence of $S_{meas}$ from the GaAs line and evaluated $S^I_{in}$ using the parameters shown in Table 1. Figure 12(a) shows the linear conductance $G$ of the QPC in units of $G_0 = 2e^2/h$ as a function of $V_{SG}$ [ohmic contact resistance ($\cong 2$ kΩ) is subtracted]. The current-noise measurements were performed at $G/G_0$ = 1.5, 1, 0.65, and 0, indicated by the markers superimposed on the trace. Figure 12(b) shows the $V_{bias}$ dependence of the bias-induced excess noise $S^I_{bias} = S^I_{in}(V_{bias}) - S^I_{in}(0)$. We observe that $S^I_{bias}$ increases with $|V_{bias}|$ at $G/G_0$ = 1.5 (filled circles) and 0.65 (filled squares), while it remains zero independent of $|V_{bias}|$ at $G/G_0$ = 1 (open diamonds) and 0 (open triangles), which is consistent with theory[2] and previous experiments[13,14,38-42]. We compare the $S^I_{bias}$ data at $G/G_0$ = 1.5 and 0.65 with theoretical shot noise

$$S_{shot} = 2e \frac{V_{bias}}{R_m} F \left[ \coth\left(\frac{eV_{bias}}{2k_B T_e}\right) - \frac{2k_B T_e}{eV_{bias}} \right], \quad (13)$$

where $F = \Sigma_{n,\sigma}[T_{n,\sigma}(1 - T_{n,\sigma})]/\Sigma_{n,\sigma} T_{n,\sigma}$ is the Fano factor. Here, $T_{n,\sigma}$ is the transmission probability of spin $\sigma = \uparrow$ or $\downarrow$ electrons in the $n$th subband in the QPC; in the present spin-degenerate case, we can assume $T_{n,\uparrow} = T_{n,\downarrow}$. The theoretical curve obtained by substituting $T_{1,\sigma} = 1$ and $T_{2,\sigma} = 0.5$ ($T_{1,\sigma}$ = 0.65 and $T_{2,\sigma} = 0$) into Eq. (13), shown in blue in the figure, agrees well with the experimental result at $G/G_0$ = 1.5 (0.65)[43].

**D. Resolution**

The resolution of the current-noise measurement was evaluated as the standard deviation of measured $S^I_{in}$[11]. We repeated the $S_{meas}$ measurement about two-hundred times with the QPC set at the first conductance plateau ($R_m \cong 15$ kΩ, including the ohmic contact resistance) and $V_{bias}$ = 0 V and then estimated $S^I_{in}$ at $f = f_1$ for each measurement. The upper panel of Fig. 13(a) [13(b)] shows the deviation $\Delta S^I_{in} = S^I_{in} - \langle S^I_{in} \rangle$, where $\langle S^I_{in} \rangle$ is the average of the $S^I_{in}$ values, measured through the GaAs (ATF) line. The lower panel is the result of the histogram analysis. The Gaussian fit for the histogram tells us that the standard deviation $\sigma_{GaAs}$ ($\sigma_{ATF}$) of the GaAs (ATF) line is $0.52 \times 10^{-29}$ A$^2$/Hz ($0.83 \times 10^{-29}$ A$^2$/Hz). When we define the resolution as $\delta S^I_{in} = \sigma_{GaAs}$ ($\sigma_{ATF}$), we find that the GaAs line has about 1.6 times better resolution than the ATF line. Note that, in the present experimental setup shown in Fig. 11(a), the GaAs and ATF lines share $S^I_{HEMT}$ (see Table 1), so the lower $\sigma_{GaAs}$ is only due to the lower $S^V_{HEMT}$ of the GaAs line. If the ATF line is removed to reduce $S^I_{HEMT}$, the GaAs line may show even better resolution.

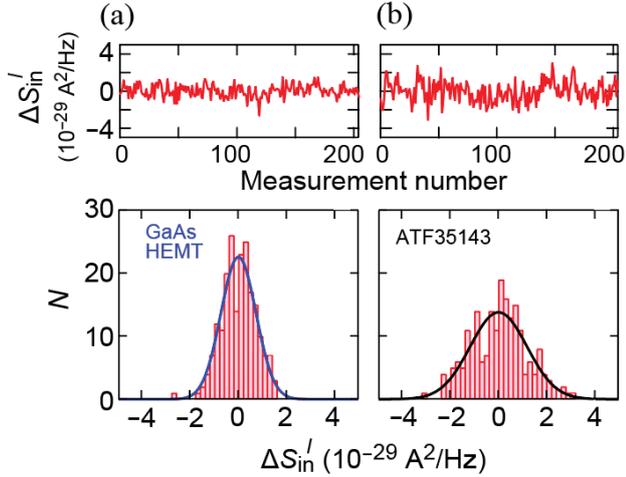

Fig. 13. (a), (b) (upper panels) $\Delta S^I_{in}$ data for the same measurements repeated 205 times ($R_m \cong 15$ k$\Omega$; $\tau_{int} = 50$ s). (lower panels) Histogram analysis. Solid lines show Gaussian fits. (a) GaAs HEMT line shows $\sigma_{GaAs} = 0.52 \times 10^{-29}$ A$^2$/Hz, while (b) ATF HEMT shows $\sigma_{ATF} = 0.83 \times 10^{-29}$ A$^2$/Hz.

It is instructive to compare our results with that in a previous study. Reference 11 reports $\delta S^I_{in} = 2.8 \times 10^{-29}$ A$^2$/Hz with $\tau_{int} = 10$ s for a cross-correlation measurement using ATF-HEMT amplifiers. If we perform a similar measurement using our GaAs-HEMT amplifiers, the standard deviation, or the resolution, is expected to be $\delta S^I_{in} = 0.82 \times 10^{-29}$ A$^2$/Hz: the shorter $\tau_{int}$ of 10 s (instead of the 50 s used in this study) would increase $\delta S^I_{in}$ by a factor of $\sqrt{5}$, while the cross-correlation technique would decrease it by $\sqrt{2}$. Thus, the resolution of our setup is less than one third of that reported in Ref. 11. If we define the measurement efficiency by the inverse of $\tau_{int}$, which is the data integration time to reach a certain value of $\delta S^I_{in}$, the efficiency of our system is more than $(1/3)^{-2} = 9$ times better than the previous one.

## VI. SUMMARY

We have presented a noise-measurement system composed of a homemade cryogenic GaAs-HEMT amplifier. Our system was precisely calibrated by Johnson-noise thermometry. The system has a higher resolution than that composed of a commercial-HEMT amplifier, mainly due to the high $g_m$ of the homemade HEMT. Finally, we add that further improvement is possible by using a HEMT with higher $g_m$, which will be achieved by enhancing the mobility and density of the 2DES and optimizing the HEMT structure.

## ACKNOWLEDGEMENTS

We appreciate technical support from T. Shimizu and H. Murofushi. This study was supported by Grants-in-Aid for Scientific Research (JP16H06009, JP19H05603, JP15H05854, JP19H05826, and JP19H00656), JST PRESTO Grant Number JP17940407, and RIEC, Tohoku University.

## DATA AVAILABILITY

The data that support the findings of this study are available from the corresponding author upon reasonable request.

[32] The saturation of $I_{ch}$ near 20 mA for the $W$ = 3 mm HEMT is due to parasitic resistance of about 25 Ω, which is the sum of the ohmic contact resistance and the wiring resistance. We do not observe such saturation for the $W$ = 1 mm HEMT over the measured $V_g$ range, because of its smaller contact resistance and the resultant smaller total parasitic resistance (about 20 Ω).
[33] The commercial HEMT becomes unstable above $V_{DD}$ = 0.4 V.
[34] We restricted the measurement up to $P \cong 1$ mW to avoid the temperature rise of the 4 K stage.
[35] The $S^V_{HEMT}$ peak near $f$ = 900 kHz is an artifact originating from the noise in the power supply.
[36] At higher $R_L$, the operating point becomes close to the pinch-off, leading to lower $g_m$. This increases $S^V_{HEMT}$ for a given $S^I_{ch}$ because $S^V_{HEMT} \cong S^I_{ch}/g_m^2$.
[37] In this simulation, we first examined fits to the data for the GaAs line with several sets of $|A(f_1)|$ and $T_e$ values to obtain $r$, $S^I_{HEMT}$, and $S^V_{HEMT}$ as fit parameters, while $C_{in}$ = 238 pF and $L_{in}$ = 33 μH were fixed to give $f_1$ = 1.794 MHz. Then, we examined fits to the data for the ATF line with the fixed values of $T_e$, $r$, and $S^I_{HEMT}$ obtained from the former fit. Using this procedure, we found that only the parameters presented in Table 1 and $r$ = 14.7 Ω explain our experimental data consistently within the error of approximately 4%.
[38] M. Reznikov, M. Heiblum, H. Shrtikman, and D. Mahalu, Temporal correlation of electrons: suppression of